\documentclass{aa}
\def \th {\thinspace}
\def \sax {{\it BeppoSAX}}
\def \degmark{^\circ}

\def \arcmin {\hbox{$^\prime$}}

\def\approxgt{\mathrel{\hbox{\rlap{\lower.55ex \hbox {$\sim$}}
\kern-.3em \raise.4ex \hbox{$>$}}}}
\def\approxlt{\mathrel{\hbox{\rlap{\lower.55ex \hbox {$\sim$}}
\kern-.3em \raise.4ex \hbox{$<$}}}}
\def \th {\thinspace }
\def \ref {\reference{}}
\def \sax {{\it BeppoSAX}}

\def \degmark{^\circ}

\def \arcmin {\hbox{$^\prime$}}

\usepackage{psfig}
\usepackage{psbox}
\usepackage{epsf}
\usepackage{graphicx}
\begin{document}

\thesaurus{6(13.25.5;  
             08.14.1;  
             08.02.1;  
             02.01.2)} 

\title{Results of a LMXB survey: variation in the height
of the neutron star blackbody emission region}

\author{M. J. Church\inst{1,2} and M. Ba\l uci\'nska-Church\inst{1,2}}
\offprints{M. J. Church}   
\institute{School of Physics and Astronomy, University of Birmingham,
                 Birmingham B15 2TT\\
                 email: mjc@star.sr.bham.ac.uk, mbc@star.sr.bham.ac.uk\\
\and                 
                 Institute of Astronomy, Jagiellonian University, ul.
Orla 171, 30-244 Cracow, Poland}

\date{Received 6 October 2000; Accepted 29 January 2001}
\authorrunning{Church and Ba\l uci\'nska-Church}
\titlerunning{Neutron star blackbody emission}
\maketitle

\begin{abstract}
We present results of a survey of the spectra of Low Mass 
X-ray Binaries using {\it ASCA}. It is shown that all sources in the survey are 
well-fitted by the same two-component emission model that we have previously shown is able 
to describe both the non-dip and dip spectra of the dipping class of LMXB. This model
consists of point-like blackbody emission from the neutron star plus Comptonized 
emission from a disk-like accretion disk corona of radius typically 50,000 km. 
Additional data from results published elsewhere by us from \sax\ and
{\it ASCA} are added to the survey. 
The large variation in blackbody luminosity of survey sources 
is shown to be due primarily to major changes 
in blackbody emitting area. Fitting a multi-temperature disk 
blackbody plus Comptonization model to the survey spectra requires values of inner disk radius 
substantially less than the neutron star radius in many cases, making disk origin of the blackbody
highly unlikely. Assuming that the emission is from an equatorial strip on the 
neutron star, it is shown that the half-height of the strip {\it h} agrees well with the half-height 
{\it H} of the radiatively-supported inner accretion disk, this agreement spanning three orders of 
magnitude in each parameter. Possible mechanisms for the agreement
are discussed, including radial accretion flow between inner disk and
star, and accretion flow ``creep'' on the surface of the neutron star.
\end{abstract}

\keywords   {X rays: stars --
             stars: neutron --
             binaries: close --
             accretion: accretion disks}

\section{Introduction}
Low mass X-ray binaries (LMXB) may be classified according to inclination angle and according 
to colour-colour properties. Frank et al. (1987) explained the markedly different 
light curves of LMXB in terms of a model of varying inclination. Sources
having inclination angles between zero and $\rm {\sim 65^{\degmark}}$ will 
show no orbital related variability, sources with inclination
of 65--$\rm {80^{\degmark}}$ will exhibit X-ray dipping, and sources with inclinations of 
80--$\rm {90^{\degmark}}$ will be 
Accretion Disk Corona (ADC) sources. Strong orbital-related
variability is seen in the light curves of the dipping and ADC sources,
but not in the other sources which are viewed from above the orbital
plane. The second type of classification is based on examining the
behaviour of LMXB on X-ray colour-colour diagrams. Hasinger et al. (1989) 
first used the term ``Z-track'' to describe the evolution on a
colour-colour diagram of bright LMXB such as \hbox {Sco\th X-1} and
\hbox {Cyg\th X-2}, the sources clearly undergoing marked spectral changes
as they moved along the horizontal, normal and
flaring branches. The Atoll sources were similarly recognised as
having different colour-colour tracks (Hasinger \& van der
Klis 1989). Associated with tracks in the Z-track and Atoll sources 
was correlated timing behaviour, especially of QPOs 
(e.g. van der Klis 1995).

Clearly, strong spectral changes take place during track movement, but 
these are not understood. A number of surveys of the spectra of the Atoll 
and Z-track sources have been made, including that based on {\it Exosat} 
data (White et al. 1988). Schulz et al. (1989) classified a large number of sources in terms of 
colour-colour diagrams, and carried out spectral fitting using a 
blackbody plus cut-off power law model. A similar blackbody plus power
law model was used by Schulz (1999) in a {\it Rosat} survey of LMXB.
Christian \& Swank (1997) produced a compilation 
of the properties of LMXB from the {\it Einstein} solid state and
monitor proportional counter detectors.

There are considerable differences between the emission
models that have been proposed for LMXB in general. A disk blackbody plus a Comptonized
blackbody model was proposed by Mitsuda et al. (1989). A model also
based on accretion disk blackbody emission 
was applied by Czerny et al. (1986) 
to {\it Einstein} data. The generalised thermal model (White et al. 1985)
was used by White et al. (1988), with a blackbody added for high-luminosity sources 
which was not required for the lower luminosity burst sources, from 
which they concluded that spectral formation was dominated by 
Comptonizaton in the inner accretion disk. More recently, a
two-component model proposed by Church and Ba\l uci\'nska-Church (1995)
has been shown to explain spectral evolution in the dipping
LMXB sources, consisting of point-source blackbody emission identified with 
the neutron star plus Comptonized emission from an extended ADC
modelled by a cut-off power law (Church \& Ba\l uci\'nska-Church 1993, 1995; 
Church et al. 1997, 1998a,b; Ba\l uci\'nska-Church et al. 1999, 2000,
2001).

For the dipping LMXB sources, spectral analysis is
more strongly constrained because of the requirement that a model fits 
not only the non-dip spectrum but also several levels of dipping.
A major advantage of investigating spectral evolution in these sources
is that dipping is characterised by the slow removal of 
Comptonized emission, and measurement of dip ingress times proves that
the Comptonizing ADC has radial extent typically 50,000 km (Church
2000), which is strongly inconsistent with Comptonization models
involving a small region in the locality of the neutron star, allowing
such models to be discounted.

\begin{table}[!hb]
\caption{The observations}
\begin{center}
\begin{tabular}{lll}
\noalign{\smallskip}
\hline
Source &Type &Date  \\
\noalign{\smallskip}
\hline\smallskip
GX\th 9+9                &Atoll          &1994, Mar 25 \cr
GX\th 13+1               &Atoll          &1994, Sep 10 \cr
4U\th 1636-536           &Atoll          &1993, Aug 9  \cr
Ser\th X-1               &Prob Atoll     &1994, Sep 20 \cr
X\th 2127+119            &Prob Atoll     &1995, May 16 \cr
Aql\th X-1               &               &1994, Apr 30 \cr
X\th 1746-371            &Dipping        &1995, Sep 21 \cr
XB\th 1254-690           &Dipping        &1994, Mar 18 \cr
GX\th 5-1                &Z-track        &1993, Oct 9 \cr
Cyg\th X-2               &Z-track        &1993, Jun 10 \cr

\noalign{\smallskip}
\hline

\end{tabular}
\end{center}
\end{table}

The success of this emission model with the dipping sources suggests that the model be tested 
with the other classes of LMXB, i.e. the Z-track and Atoll sources specifically. 
Other authors have successfully applied two-component blackbody + Comptonization models 
to Z-track and Atoll sources (above, and e.g. Hasinger et al. 1990, Barret et al. 2000). 
However, these often ascribe emission regions to the blackbody and Comptonized emission 
such as the accretion disk and a localised inner region respectively, 
e.g. Barret et al. (2000), and so physically, these models differ radically from ours
and are also inconsistent with our measured ADC sizes. Recently, the
{\it comptt} implementation of the Titarchuk (1994) Comptonization model
has been used (together with a blackbody term) to fit {\it BeppoSAX} data on LMXB 
sources (e.g. in't Zand et al. 1999). It has been suggested that {\it
comptt} is preferable to the cut-off power law which may overestimate
the spectrum at low energies because of lack of soft seed photons.
However, it is easy to calculate the spectrum of the disk 
(expressed as the photon number flux per keV) integrated to
a radius equal to our measured ADC sizes, by combining the temperature
gradient $T(r)$ from standard thin disk theory with Planck's equation.
This shows that for typical LMXB luminosities, the integrated spectrum
forms a broad peak between 0.001 and 0.1 keV, thus producing a huge
sea of soft photons, so that the cut-off power law is perfectly good
above $\sim $0.1 keV. However, fitting the {\it comptt} model
has produced high values of average $kT$ for the seed photons, e.g.
1.0--2.0 keV (Guainazzi et al. 1998) which are $>>$ than the above expected values
and inconsistent with the assumption of Titarchuk (1994) that the seed photons are very soft: 
$kT$ $<<$ $kT_{\rm {e}}$ (the electron temperature).
This will lead to a substantial underestimation of the Comptonized
spectrum at $\sim $1 keV as the Wien approximation (good for $h\nu $
$>>$ kT) is used in {\it comptt} to describe the seed photon spectrum.

In this paper, we apply the blackbody plus cut-off power law 
model to as many non-dipping Atoll and Z-track 
sources observed using {\it ASCA} as possible plus the non-dip spectra
of two dipping sources. We show that the model provides a good 
description of the spectra  of these sources.

\begin{figure*}[!ht]   
\begin{center}
\epsfxsize=160 mm
\leavevmode
\epsffile{f1a.eps}
\vskip 1mm
\epsfxsize=160 mm
\leavevmode
\epsffile{f1b.eps}
\vskip 1 mm
\epsfxsize=160 mm
\leavevmode
\epsffile{f1c.eps}
\caption{Best fits to the {\it ASCA} spectra of 10 LMXB using the two-component 
continuum model discussed in the text, either in the form of blackbody + power 
law or blackbody + cut-off power law. Lines were added to the model in
the 4 sources where definite line detections were made. In all 
cases the ordinate is in units of photon cm$^{-2}$ s$^{-1}$ keV$^{-1}$
\label{fig1}}
\end{center}
\end{figure*}

\section{Analysis and Results}

The observations of 10 LMXB with {\it ASCA} (Tanaka et al. 1994) 
were made between 1993, June and 1995, September
as shown in Table 1. The designation ``probably atoll'' was taken from
van der Klis (1995) when the type is not well known.
For each source, raw data files containing less than 100 counts were deleted, and 
the temperature-based gain correction was made. Screening was
carried out after inspection of plots of the housekeeping parameters.
Data were selected with the source less than
$\rm {0.01^{\degmark}}$ from the telescope pointing direction, elevation 
above the Earth's rim more than $\rm {5^{\degmark}}$ and geomagnetic rigidity
more than 6 GeV/c. Particle background was rejected for values of
the background monitor rate of more than 200 count s$^{-1}$ in most cases.
Source data were selected from a 6\arcmin\ radius circular region
in the image. Many of the sources were bright and so background subtraction 
was not important; when necessary, background data were extracted from an
appropriate 6\arcmin\ region. Deadtime correction was performed on
both light curves and spectra; this enabled high and medium bit rate data 
to be combined in a single light curve. In all cases except one, GIS2
data were used; in the case of X\th 1746-371,
GIS3 data were used because of problems with the GIS2 raw data.

Light curves were extracted in the total GIS energy band
and also in 3 standard bands: 0.7--4.0 keV, 4.0--7.0 keV and 
7.0--10.0 keV. These were used to examine source variability 
and to construct colour-colour diagrams using 
the ratio of count rates 7.0--10.0 keV/4.0--7.0 keV 
and 4.0--7.0 keV/0.7--4.0 keV.
Generally, there was little evidence for movement on the colour-colour
diagrams during observations; some of the sources were constant,
some showed slow drifts in intensity and some more rapid
variability. To ensure that data could not be mixed from different 
intensity states, spectra were selected from appropriately
narrow intensity bands for all sources, for example, in the band 75--80 
count s$^{-1}$ for GX\th 13+1. Response files were 
constructed appropriate to the position of each source in the image, and
the latest corrections applied to the responses to correct for known
differences between the detectors. 

Spectra were grouped to a minimum of 100 counts 
per bin for strong sources and to 20 counts per bin for faint sources, 
and systematic errors of 2\% added (Fukazawa et al. 1997). For each 
source, a range of models was tried including simple one-component models 
such as absorbed power law  ({\sc ab*pl}, where {\sc ab} is the absorption term and 
{\sc pl} the power law), absorbed
\begin{table*}
\caption{Spectral fitting results. In most cases, the blackbody +
cut-off power law model did not give any improvement over the
blackbody + power law model, indicating a cut-off energy much higher
than 10 keV. Only in cases where there was an improvement are results
for both of these models given.\label{2}}
\begin{center}
\begin{tabular}{llllrrr}
\noalign{\smallskip}
\hline
Source &Model &\hfil $N_{\rm {H}}$ \hfil &$kT$  &\hfil $\Gamma$ \hfil
&$E_{\rm {CO}}$  &$\chi^{\rm {2}}$/dof \\
&&$\rm {10^{22}}$ cm$^{-2}$     &kev&&keV\\
\noalign{\smallskip}
\noalign{\smallskip}
\hline\smallskip
Atoll sources \\
\noalign{\smallskip}
\hline\smallskip
GX\thinspace 9+9 &Power law & $0.55\pm 0.01$ & \dots & $1.89\pm 0.07$&\dots &4171/712  \\
&Cut-off power law & $0.19\pm 0.01$ & \dots &$ 0.80\pm 0.03$&$4.1\pm 0.1 $ & 507/711\\
&Blackbody + power law & $0.37\pm 0.02$ & $1.17\pm 0.02$ &$1.95\pm 0.03$& \dots &615/710\\
&Blackbody + cut-off power law &$0.26\pm 0.01$ & $1.27\pm 0.06$&$1.21\pm 0.03$ & $\sim $6&514/710\\

\hline
GX\th 13+1 &Power law & $4.4\pm 0.1$ & \dots & $2.51\pm 0.02$ &\dots&1639/502  \\
&Cut-off power law & $2.4\pm 0.1$ & \dots &$ -0.30\pm 0.13$&$1.9\pm 0.1 $ & 459/501\\
&Blackbody + power law & $3.5\pm 0.2$ & $1.18\pm 0.02$ &$2.88\pm 0.22$&\dots &446/499\\

\hline
4U\th 1636-536 &Power law & $0.55\pm 0.02$ & \dots & $2.12\pm 0.02$&\dots &423/396  \\
&Cut-off power law & $0.46\pm 0.03$ & \dots &$ 1.83\pm 0.09$&$13.5^{+6.2}_{-3.2} $ & 395/395\\
&Blackbody + power law & $0.43\pm 0.05$&$0.65^{+0.08}_{-0.05}$&$1.99\pm 0.07$& \dots &391/394\\

\hline
Ser\thinspace X-1 &Power law & $0.71\pm 0.02$ & \dots & $1.85\pm 0.02$& \dots &1253/605  \\
&Cut-off power law & $0.43\pm 0.02$ & \dots &$ 1.05\pm 0.05$&$5.4\pm 0.4 $ & 513/604\\
&Blackbody + power law & $0.62\pm 0.03$ & $1.26\pm 0.08$ &$2.00\pm 0.05$& \dots &514/603\\

\hline
X\th 2127+119 &Power law & $0.58\pm 0.04$ & \dots & $2.04\pm 0.04$&\dots &400/211  \\
&Cut-off power law & $0.14\pm 0.06$ & \dots &$ 0.56\pm 0.19$&$2.7^{+0.4}_{-0.3} $ & 218/210\\
&Blackbody + power law & $0.24\pm 0.09$ & $0.89\pm 0.06$ &$1.85\pm 0.16$& \dots &216/209\\

\hline
Aql X-1 &Power law & $0.64\pm 0.03$ & \dots & $1.75\pm 0.02$&\dots&535/481  \\
&Cut-off power law & $0.48\pm 0.05$ & \dots &$ 1.39\pm 0.10$&$12.4^{+4.2}_{-2.6} $ & 497/480\\
&Blackbody + power law & $0.49\pm 0.08$&$0.92^{+0.25}_{-0.10}$&$1.66\pm 0.08$& \dots &503/479\\

\noalign{\smallskip}
\hline\smallskip
Dipping sources \\
\noalign{\smallskip}
\hline\smallskip
XB\th 1746-371 & Power law & $0.50\pm 0.05$ & \dots & $1.83\pm 0.06$&\dots &129/106  \\
&Cut-off power law & $0.31\pm 0.09$ & \dots &$ 1.20\pm 0.27$&$6.4^{+4.9}_{-2.0} $ & 115/105\\
&Blackbody + power law & $0.50\pm 0.05$&$1.22^{+0.31}_{-0.21}$&$1.91\pm 0.05$& \dots &283/276\\

\hline
XB\th 1254-690 & Power law & $0.48\pm 0.03$ & \dots & $1.83\pm 0.03$&\dots &389/281  \\
&Cut-off power law & $0.21\pm 0.05$ & \dots &$ 0.93\pm 0.13$&$4.6^{+0.8}_{-0.6}$ & 259/278\\
&Blackbody + power law & $0.38\pm 0.06$ & $1.17\pm 0.11$ &$1.96\pm 0.13$&\dots &264/277\\

\noalign{\smallskip}
\hline\smallskip
Z-track sources\\
\noalign{\smallskip}
\hline\smallskip
GX\thinspace 5-1 &Power-law & $4.8\pm 0.1$ & \dots & $2.56\pm 0.03$&\dots &952/438  \\
&Cut-off power law & $3.0\pm 0.1$ & \dots &$ 0.20\pm 0.18$&$2.2\pm 0.2$ & 476/437\\
&Blackbody + power law & $3.4\pm 0.2$ & $1.05\pm 0.04$ &$2.29\pm 0.18$& \dots &475/436\\

\hline
Cyg\thinspace X-2 &Power-law & $0.56\pm 0.01$ & \dots & $2.11\pm 0.01$& \dots &3186/541  \\
&Cut-off power law & $0.06\pm 0.02$ & \dots &$ 0.45\pm 0.05$&$2.5\pm 0.1 $ & 483/540\\
&Blackbody + power law & $0.29\pm 0.02$ & $1.01\pm 0.02$ &$2.16\pm 0.05$& \dots &559/539\\
&Blackbody + cut-off power law & $0.10\pm 0.02$ & $1.06\pm 0.12$&$0.75\pm 0.06$ & $\sim $3
&487/539\\

\noalign{\smallskip}
\hline
\end{tabular}
\end{center}
\end{table*}
\noindent blackbody ({\sc ab*bb}) and absorbed bremsstrahlung ({\sc
ab*br}). The absorbed 
blackbody model always gave totally unacceptable
fits with $\chi ^2$/degree of freedom (dof) typically 10; for the
bremsstrahlung models the fits were also not acceptable with
$\chi ^2$/dof varying between 1.2 and 2.0. 
Two-component models consisting
of blackbody plus a Comptonized component were tried, in the forms
{\sc ab*(bb + pl}) and {\sc ab*(bb + cpl}). The first form was used to approximate
a cut-off power law at energies well below the Comptonization break.
In several cases discussed below, it was necessary to add lines
to the continuum model to obtain an acceptable fit. Results for the 
absorbed power law, absorbed cut-off power law and two-component
models are given in Table 2 and these were obtained with lines
added if appropriate.

In most sources, there was clear curvature, i.e. downcurving, in the 
spectrum between 1 and 10 keV (see Fig. 1), such that the {\sc ab*pl} model gave poor fits
as can be seen from the $\chi ^2$ values in Table 2. The generalised thermal
(model {\sc ab*cpl}, where {\sc cpl} is a cut-off power law),
provided good fits to all of the spectra. However, in the cases of 
several sources, particularly GX\th 13+1, X\th 2127+119, GX\th 5-1 and 
Cyg\th X-2, this was only achieved with very small and even negative values of 
the power law index $\Gamma $ which may be regarded as improbable. The
values of cut-off energy $E_{\rm {CO}}$ are all very small 
($\sim $few keV) which is typical
of this model. However, we know from work on several dipping sources using \sax\
that the break energy is often much greater than 10 keV (Church et al.
1998b; Ba\l uci\'nska-Church et al. 1999, 2000)
and so the low values in most cases are unlikely to be real, but result
from this model fitting the curvature between 1 and 10 keV
actually due to the presence of a blackbody. Fitting a
two-component model showed that in all cases, either the blackbody plus power law
model or the blackbody plus cut-off power law model gave a fit as good
or better than the {\sc ab*cpl} model, but without requiring any odd 
values. In most cases, the simpler {\sc ab*(bb + pl}) model gave 
a fit equally as good as the {\sc ab*(bb + cpl}) model showing that
$E_{\rm{CO}}$ was $>>$ 10 keV and so not possible to be determined using {\it ASCA}. 
In the case of GX\th 9+9 and Cyg\th X-2 this was not the case, showing
that $E_{\rm {CO}}$ was less than 10 keV. Fitting for these sources, 
was carried out with $E_{\rm {CO}}$ fixed at a sequence of values as discussed below.

\begin{table}
\caption{Line detections\label{3}}
\begin{center}
\begin{tabular}{lllr}
\noalign{\smallskip}
\hline
Source &Energy &Width $\sigma $ &EW  \\
        &keV  &keV        & eV\\
\noalign{\smallskip}
\hline\smallskip
GX\th 13+1               &$6.37\pm 0.23$ &0.1     &44\cr
Ser\th X-1               &6.6 &0.17&81\cr
Aql\th X-1               &$0.99\pm 0.04$ &0.01    &53\cr
                         &$6.81\pm 0.16$ &0.001   &43\cr
XB\th 1254-690           &$6.40\pm 0.12$ &0.001   &98\cr
\noalign {\smallskip}
\hline
\end{tabular}
\end{center}
\end{table}

Lines were detected in the sources GX\th 13+1, Aql\th X-1, Ser\th X-1
and XB\th 1254-690, and the energies, widths and equivalent widths are shown in Table 3.
These agree generally with the detections made in the extensive study of
lines in LMXB using {\it ASCA} by Asai et al. (2000).
There was also evidence for a weak line in XB\th 1746-371 at $\sim $6.4 keV
also detected in SIS by Asai et al., and a possible weak detection 
of a line at $\sim $1.5 keV in GX\th 13+1. In Ser\th X-1, the iron
feature detected was broad, and the width was not well-constrained using
GIS data; consequently, to obtain the best continuum fit,
the line energy and width $\sigma $ were set to the values
determined by Asai et al. The best-fit two-component model
for each source is shown in Fig. 1.

\begin{table*}[!ht]
\caption{Results of fitting a disk blackbody + cut-off power law model to {\it ASCA} and 
{\it BeppoSAX} data. The predominance of small or negative values of photon index $\Gamma $ 
can be seen, and the
unphysical values of $r_{\rm {i}}$ less than the radius of the neutron star.
\label{4}}
\begin{center}
\begin{tabular}{lllllll}
\noalign{\smallskip}
\hline
Source &Instrument &$kT_{\rm {i}}$ & $\Gamma $& $E_{\rm {CO}}$&$\chi^{\rm {2}}$/dof&$r_{\rm {i}}$\\
                && keV   &&keV   &&km\\
\noalign{\smallskip}
\hline\smallskip
GX\th 9+9     &ASCA GIS &0.70$\pm $0.10&-0.01$^{+0.15}_{-0.60}$&2.7$^{+0.2}_{-0.4}$&490/709&10--17$^{+3}_{-2}$\cr
GX\th 13+1    &ASCA GIS &1.14$^{+0.32}_{-0.23}$ &-3.00$\pm $0.01&1.17$\pm $0.01&435/499&11--19$^{+10}_{-6}$\cr
4U\th 1636-536&ASCA GIS &0.64$\pm $0.07&+0.50$^{+0.17}_{-0.79}$&4.0$\pm $0.8&365/393&14--24$^{+7}_{-4}$\cr
Ser\th X-1   &ASCA GIS &2.42$^{+0.13}_{-0.08}$&+0.20$^{+0.01}_{-0.03}$&0.85$^{+0.80}_{-0.34}$&489/601&2.7--4.7$^{+0.3}_{-0.6}$\cr
X\th 2127+119 &ASCA GIS &3.22$\pm $1.24&-0.04$^{+0.40}_{-0.86}$&1.5$^{+0.2}_{-0.8}$&213/208&0.79--1.1$^{+3.0}_{-1.1}$\cr
Aql\th X-1    &ASCA GIS &1.00$\pm $0.21&-1.05$\pm $1.62&2.8$^{+3.3}_{-0.8}$&482/480&2.8--4.8$^{+1.2}_{-0.8}$\cr
XB\th 1746-371&ASCA GIS &3.02$^{+3.06}_{-0.83}$&-1.14$^{+0.58}_{-1.67}$&0.85$\pm $0.01&112/103&1.0--1.4$^{+1.5}_{-1.1}$\cr
XB\th 1254-690&ASCA GIS &2.51$^{+0.50}_{-1.93}$&+1.42$\pm $0.63&3.98$\pm $0.1&257/276&2.0--2.9$\pm $2.1\cr
GX\th 5-1     &ASCA GIS &1.65$^{+0.06}_{-0.13}$&-1.36$\pm $0.10&9.8$^{+3.0}_{-2.0}$&473/435&15--26$^{+4}_{-2}$\cr
Cyg\th X-2    &ASCA GIS &1.83$\pm $0.17&-0.90$\pm $0.04&0.74$\pm $0.01&480/538&8.6--9.8$^{+1.0}_{-3.0}$\cr        
XB\th 1916-053&SAX NFI  &2.96$\pm $0.24&+1.60$\pm $0.04&200$\pm $150&556/520&0.48--0.67$\pm $0.13\cr        
XB\th 1323-619&SAX NFI &2.34$^{+0.56}_{-0.25}$&+1.17$^{+0.24}_{-0.39}$&32$\pm $11&204/216&0.48--0.67$\pm $0.01\cr
X\th 1624-490 &SAX NFI  &5.1$^{+1.6}_{-1.0}$&-0.78$\pm $0.58&1.7$\pm $0.3&291/294&0.39--0.41$^{+0.38}_{-0.21}$\cr
\noalign {\smallskip}
\hline
\end{tabular}
\end{center}
\end{table*}

\subsection{X\th 1746-371, XB\th 1254-690}
Spectral analysis results of the {\it ASCA} observations of these
dipping sources have not previously been presented. 
These sources were well fitted by the two-component model. In XB\th 1746-371, 
the best-fit results were obtained by simultaneous fitting of
the non-dip spectrum and the deepest dip spectrum. This technique has previously been
used to constrain emission parameters better in dipping sources (e.g.
Ba\l uci\'nska-Church et al. 1999).

\subsection{GX\th 9+9, Cyg X-2}

In GX\th 9+9 and Cyg\th X-2 the blackbody plus power law model gave a substantially
worse fit than the single component cut-off power law model showing
that $E_{\rm {CO}}$ is not $>>$ 10 keV. The model {\sc ab*(bb + cpl})
gave improved qualities of fit. It was difficult to determine the cut-off energy
because of the competition in modelling the  1--10 keV curvature between the blackbody and the 
Comptonization down-curving, and so $E_{\rm {CO}}$ was fixed at a series of values:
10, 8, 6, 4 and 2 keV. In GX\th 9+9, $\chi^2$ became as good as for 
the cut-off power law model for $E_{\rm {CO}}$ $\sim$6 keV. For smaller cut-off energies, $\chi^2$
continued to improve, however the power law index decreased to $<$ 1
which is unlikely to be real. For Cyg\th X-2, stepping in $E_{\rm {CO}}$ 
gave acceptable $\chi^2$ for $E_{\rm {CO}}$ $\sim $3 keV. This cut-off energy agrees
well with the value of 3.32 keV obtained by Smale et al. (1993) from {\it BBXRT}.
Thus, our preferred values of cut-off energy are $\sim $6 keV and $\sim $3  keV in GX\th 9+9 and 
Cyg\th X-2 respectively. Clearly, these should be re-measured in a wider band. However,
test showed that the results presented below are not sensitive to the
value of $E_{\rm {CO}}$.
 
\subsection{Disk blackbody modelling}
Finally, we tested the model consisting of disk blackbody emission plus a Comptonization term 
in the form of a cut-off power law, adding lines as before to the sources requiring lines. 
The 10 sources studied in the present {\it ASCA} survey were fitted, plus 3 sources recently 
studied using {\it BeppoSAX} (Ba\l ucinska-Church et al. 1999, 2000; Church et al. 1998b). Results 
shown in Table 4 include values of inner disk radius $r_{\rm {i}}$ obtained from the normalization 
of the blackbody {\it via}: $r_{\rm {i}}$ 
= $({\rm norm}\, / {\rm cos}\, i)^{\rm {1/2}}\cdot d$, where {\it i} is the inclination
angle of the binary system and {\it d} is the source distance. Inclination angles were assumed of 
0 -- $\rm {70^{\degmark}}$ for Atoll and Z-track sources, 70 -- $\rm
{80^{\degmark}}$ for dipping sources,
except for X\th 1624-490 where 60 -- $\rm {65^{\degmark}}$ was assumed, the upper
limit being set by the absence of X-ray eclipses.
For Cyg\th X-2, we use the range 58.5 -- $\rm {66.5^{\degmark}}$ (Orosz \& Kuulkers 1999).
Thus for each source a range of $r_{\rm {i}}$ values is shown, with 90\% confidence 
uncertainties derived from the normalization uncertainties shown for the larger value in each case.
Although good or acceptable fits were obtained, the results in many cases had improbably small 
or negative values of power law index. The values of $r_{\rm {i}}$ were in 8 out of 13 cases 
unphysical, i.e. $<$ the radius of the neutron star $R_{\rm {*}}$, in 5 of
these, $r_{\rm {i}}$ being 10 times smaller than $R_{\rm {*}}$.
In 5 other cases, GX\th 9+9, GX\th 13+1, GX\th 5-1, Cyg\th X-2 and 4U\th 1636-536, $r_{\rm {i}}$ was
not clearly unphysical, but the power law index was small or negative. 
We also note a tendency using {\it ASCA} to underestimate $\rm {kT_i}$
and overestimate $r_{\rm {i}}$ 
compared with {\it BeppoSAX} as shown by restricting the {\it
BeppoSAX} band to 1--10 keV, so 
in some cases, the $r_{\rm {i}}$ values in Table 4 may be too large. 
It is known that the multi-temperature disk blackbody model used can
underestimate $r_{\rm {i}}$, however, maximum correction factors of
$\sim $2--4 may be calculated based on the published work of Merloni et
al. (2000) and Kubota et al. (2000). Allowing for this, the very
small values of $r_{\rm {i}}$ in Table 4 make it unlikely that the blackbody
emission is from the disk.

\medskip
We have shown that the two-component model 
\begin{figure*}[!ht]   
\includegraphics[width=100mm,height=60mm,angle=0]{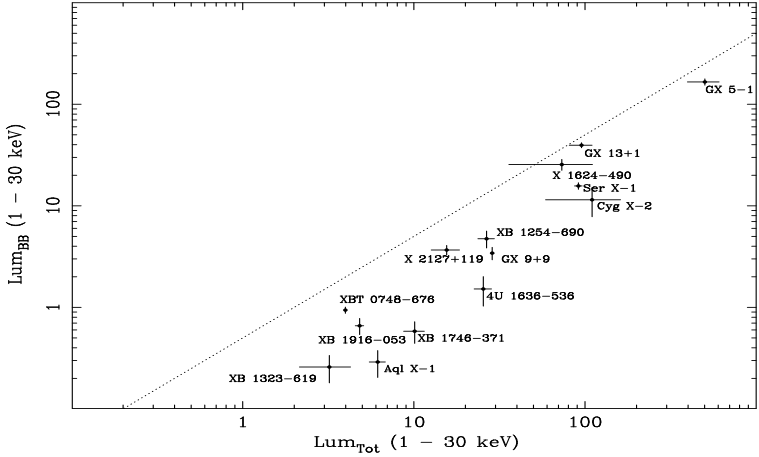}
\caption{Variation of blackbody luminosity in the band 1--30 keV with
total luminosity in the same band; both luminosities are plotted in
units of 10$^{36}$ erg s$^{-1}$. The dotted line shows the relationship expected from
simple energy considerations that the blackbody luminosity of the neutron star
should equal the luminosity of the accretion disk; i.e. 50\% of the total
luminosity\label{fig2}}
\end{figure*}
(either as {\sc bb + pl} or as {\sc bb + cpl}) gives very good fits to all of the sources in our
survey. In all cases, a blackbody component is required, and we next examine the
properties of this component.

\medskip
\subsection{The Blackbody-Total Luminosity Relation}

In Fig. 2 the variation of blackbody luminosity $L_{\rm {BB}}$ in the
band 1 -- 30 keV with the total luminosity $L_{\rm {Tot}}$ in the same band 
is shown. Also plotted in Fig. 2 are additional points from our analyses of the
\sax\ observations of XB\th 1916-053 (Church et al. 1998b), XB\th
1323-619 (Ba\l uci\'nska-Church et al. 1999) and X\th 1624-480
(Ba\l uci\'nska-Church et al. 2000)
and for XBT\th 0748-676 from {\it ASCA} data (Church et al. 1998a). These points 
are useful because the sources are weaker, and spectral parameters well-determined 
from the broadband \sax\ data. Source distances were taken from the {\it Einstein 
Observatory} survey of LMXB of Christian \& Swank (1997). In the case of XB\th 1323-619, 
a distance of 10 -- 20 kpc was derived by Parmar et al. (1989)
based on the measured peak burst flux assumed to be Eddington-limited.
As this may have underestimated the peak flux, we have
assumed a value of 10 kpc. We also assumed 10 kpc for XBT\th 0748-676,
the distance of which is not known. The sources in the present survey cover the
luminosity range $\rm {\sim 5\times 10^{36}}$ to $\rm {5\times 10^{38}}$ erg s$^{-1}$.
Errors in $L_{\rm {BB}}$ are derived from the 90\% confidence limits in the 
blackbody normalization; similarly the errors in $L_{\rm {Tot}}$  were obtained from
the 90\% errors in power law normalization. Possible errors in source distance 
will move points equal amounts on both axes.

Additionally, a line is drawn showing the Newtonian value of 50\% of
$L_{\rm {Tot}}$ that should remain at the inner disk and be available
for neutron star emission. 
In fact, the ratio of the energy available to the star to the energy in
the disk depends on the spin of the neutron star (Sunyaev \& Shakura
1986), varying from 2.2 for a slowly spinning star to small values
for a star spinning rapidly, implying in the first case that the star 
can receive 69\% of the energy. It can be seen that there is systematic 
behaviour in Fig. 2 such
that the brighter sources approach the 50\% line whereas the weaker
sources fall increasingly below this line. Schulz et al. (1989) and
Hasinger et al. (1990) found a similar variation of the blackbody luminosity with 
the total luminosity in Cyg\th X-2. It should be stressed that we do not
claim a simple relationship between $L_{\rm {BB}}$ and $L_{\rm {Tot}}$,
but rather a broad band in the Figure where the sources
lie. 
We next examine the reasons for the variation in blackbody luminosity over 
the sources investigated. 

\subsection {Blackbody Properties}

Table 5 lists the blackbody parameters derived from spectral fitting.
Included here are the blackbody temperature $kT_{\rm {BB}}$, the
blackbody luminosity in the band 1 -- 30 keV and
the blackbody radius $R_{\rm {BB}}$ defined by the relation 
$L_{\rm {BB}}$ = $4\,\pi\,R_{\rm {BB}}^{\rm {2}}\,\sigma\,T^{\rm {4}}$, where $\sigma $ is Stefan's
constant. Also shown is the half-height {\it h} of the emission region assumed to be an
equatorial strip on the neutron star. It can be seen that $L_{\rm {BB}}$ 
is reasonably well correlated with $R_{\rm {BB}}$. In Fig. 3 we show the 
variation of blackbody temperature with blackbody radius.
It is striking that the major cause of variation is the emitting area.
As the $L_{\rm {Tot}}$ increases as we move from the the extreme left of Fig. 3 
(XB\th 1323-619) to the extreme right (GX\th 5-1), $kT_{\rm {BB}}$ falls by 40\%, 
so that $T^{\rm {4}}$ decreases by a factor of 8. $R_{\rm {BB}}$ 
however, increases by a factor of $\sim $75 and emitting area by a factor of
5400 so that $L_{\rm {BB}}$ increases by a factor of $\sim $700,
consistent with the change of $L_{\rm {BB}}$ between these 
sources in Fig. 2.

\begin{table}[!ht]
\caption{Blackbody properties. {\it h} is the half-height of the
emission region on the neutron star; the blackbody luminosity
$L^{\rm {36}}_{\rm {BB}}$ (1--30 keV) is given in units of 10$^{36}$ erg
s$^{-1}$.\label{4}}
\begin{center}
\begin{tabular}{llrrrr}
\noalign{\smallskip}
\hline
Source &$d$ &  $kT_{\rm {BB}}$ & $L^{\rm {36}}_{\rm {BB}}$ & $R_{\rm {BB}}$&$h$\\
        & kpc         &keV&   & km  &km\\
\noalign{\smallskip}
\hline\smallskip
GX\th 9+9       &5.0&1.27   &1.66 &3.2&1.03\cr
GX\th 13+1      &7.0&1.18   &39.5 &12.7    &16.0\cr
4U\th 1636-536  &6.5&0.65   &1.52 &8.6     &7.5\cr
Ser\th X-1      &8.4&1.26   &15.7 &7.0     &4.9\cr
X\th 2127+119   &13.0&0.89  &3.67 &7.0     &4.9\cr
Aql\th X-1      &4.8&0.92   &0.29 &1.81    &0.33\cr
XB\th 1746-371  &9.0&1.22   &0.58 &1.4     &0.21\cr
XB\th 1254-690  &12.0&1.17  &4.74 &4.5     &2.0\cr
GX\th 5-1       &9.0&1.05  &166.0 &33.2    &110.0\cr
Cyg\th X-2      &8.0&1.06   &11.5 &8.5     &7.3\cr
XB\th 1916-053  &10.0&1.62  &0.66 &0.87    &0.08\cr
XB\th 1323-619  &10.0&1.77  &0.26 &0.45    &0.02\cr
X\th 1624-490   &15.0&1.31  &25.5 &8.3     &6.8\cr
XBT\th 0748-676 &10.0&1.99  &0.94 &0.7     &0.05\cr
\noalign {\smallskip}
\hline
\end{tabular}
\end{center}
\end{table}

Thus, the large variation in emitting area is responsible for
the large change in blackbody level. At low luminosity, the emission is
from an equatorial strip on the star, which increases in height
till at the higher luminosities, all of the star is emitting. 

\begin{figure}[!ht]    
\epsfxsize=80 mm
\begin{center}
\leavevmode \epsffile{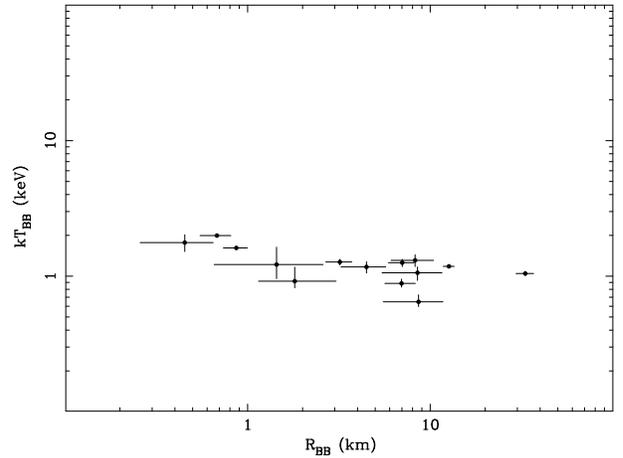}
\end{center}
\caption{Variation of blackbody temperature $kT_{\rm {BB}}$ and 
radius $R_{\rm {BB}}$ for the survey sources plus
additional sources discussed in the text. 90\% confidence limits are
shown \label{fig3}}
\end{figure}

\subsection{Accretion Disk Properties}

We next express the blackbody emitting area in terms of the
half-height {\it h} of the emitting strip assumed to be an equatorial
strip of varying height. The area is $4\,\pi\,h\,R_{\rm {*}}$ (for a
sphere intersected by two parallel planes) equal to 
$4\,\pi\,R_{\rm {BB}}^{\rm {2}}$. Thus, the half-height {\it h} 
is related to $R_{\rm {BB}}$ by

\begin{figure*}[tbh]     
\begin{center}
\parbox{8.2cm}{\psbox[size=8.0cm,rotate=r]{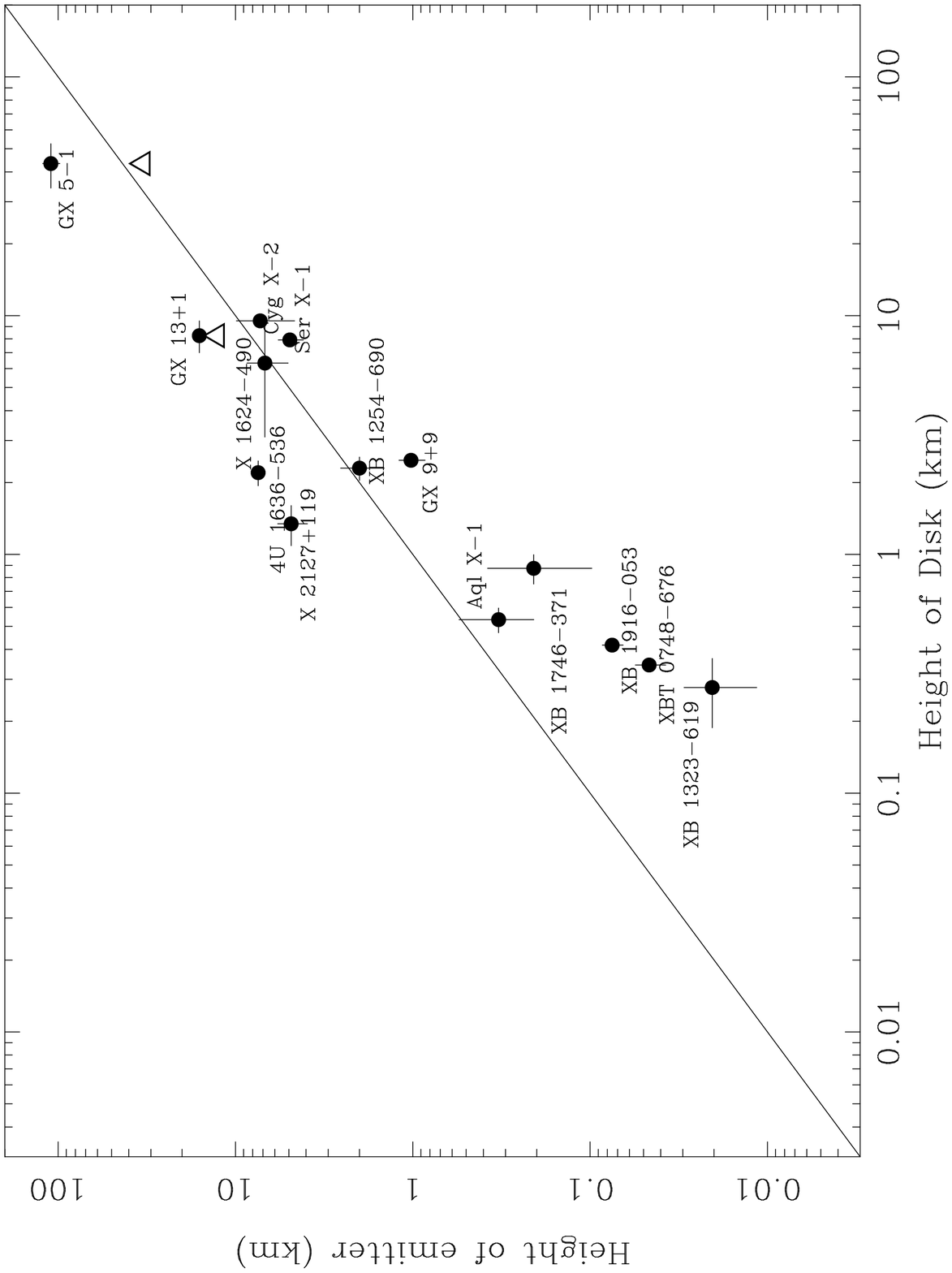}}
\parbox{8.2cm}{\psbox[size=8.0cm,rotate=r]{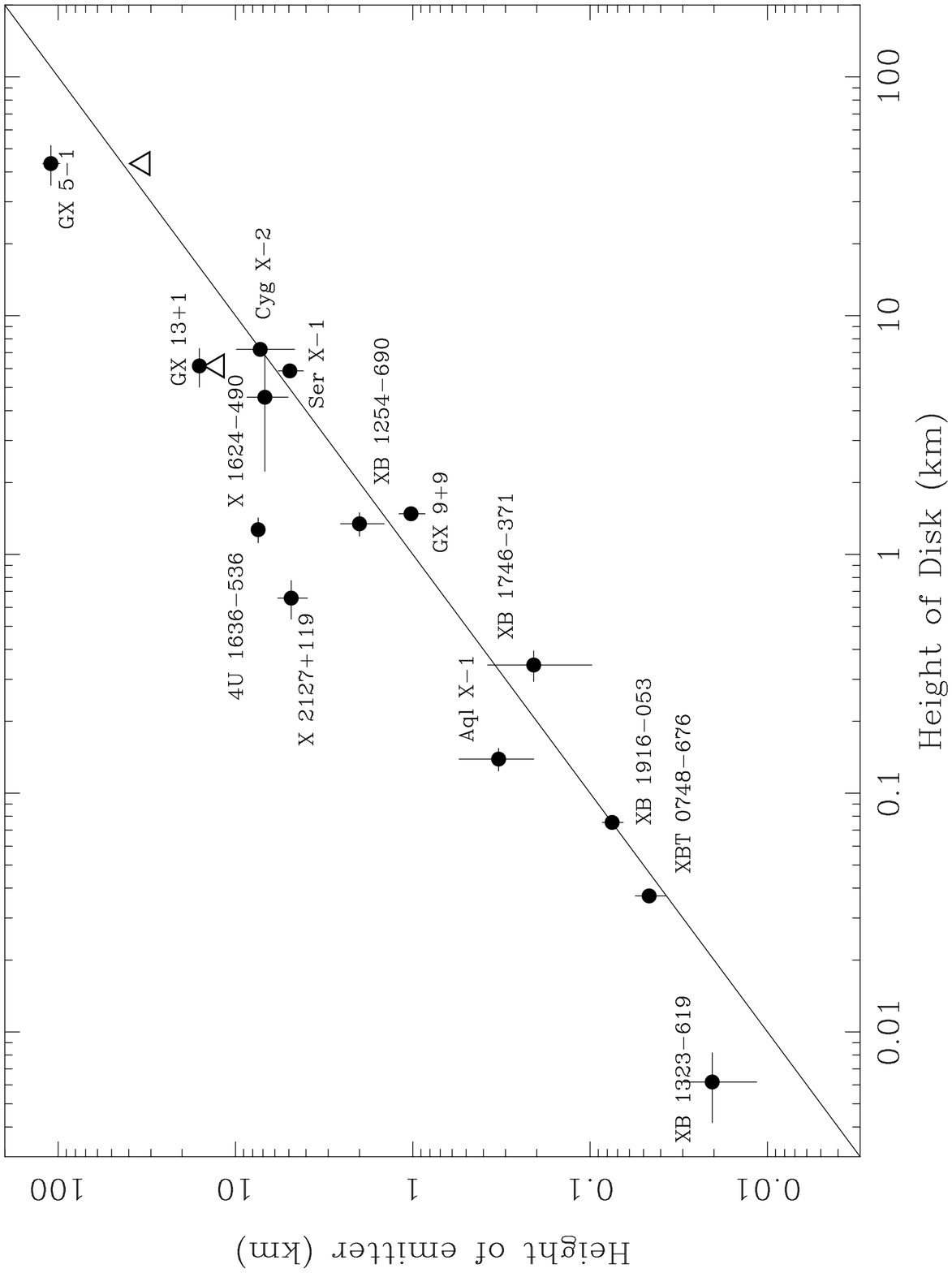}}
\caption{Left: variation of the height {\it h} of the blackbody emission region on
the surface of the neutron star with the equilibrium height $H_{\rm {eq}}$ of the radiatively
supported inner accretion disk. The triangles show the blackbody radius
$R_{\rm {BB}}$ instead of {\it h} for the two bright sources GX\th 5-1 and
GX\th 13+1. Right: {\it h} of the blackbody emission region {\it versus} the
more accurate accretion disk height {\it H} (see text)
\label{fig4}}
\end{center}
\end{figure*}

\begin{equation}
h = R_{\rm {BB}}^{\rm {2}}/R_{\rm {*}}.
\end{equation}
Values of {\it h} obtained in this way are included in Table 5.

Next, we compare {\it h} with the half-height of the inner radiatively-supported
accretion disk. For a wide range of mass accretion rates, the inner
disk is radiatively supported with half-height {\it H} given by

\begin{equation}
H = {3\,\sigma_{\rm {T}}\,\dot M\over 8\, \pi\, m_{\rm {p}}\, c}\biggl[1\,-\,
\biggl({R_{\rm {*}}\over r} \biggr)^{\rm {1/2}} \biggr]  = H_{\rm
{eq}}\biggl[1\,-\,\biggl({R_{\rm {*}}\over r} \biggr)^{\rm {1/2}}
\biggr] 
\end{equation}

\noindent where $\sigma _{\rm {T}}$ is the Thomson cross section,
$\dot M$ is the mass accretion rate and $m_{\rm {p}}$
is the proton mass (Frank, King \& Raine 1992). The disk
height {\it H} is zero at the surface of the star, but increases rapidly
until it becomes independent of $r$ having the equilibrium value $H_{\rm {eq}}$. 
Most of the increase takes place between r = 10 and r = 20 km, i.e. within 10 km 
of the stellar surface. In weak sources, with $L$ $\sim 5\times 10^{\rm {36}}$
erg $\rm {s^{-1}}$, radiative support is over a limited radial extent
so that $H_{\rm {eq}}$ is never achieved. In bright sources, $H_{\rm
{eq}}$ is achieved, and radiative support continues typically to a
radial distance of 400 km (for $L_{\rm {Tot}}$ = $\rm {10^{38}}$ erg
$\rm {s^{-1}}$). Values of $H_{\rm {eq}}$ are, of course, much larger
than the height of the gas-dominated thin disk calculated using
the solution of Shakura \& Sunyaev (1976) which for luminosities between 
10$^{36}$ and 10$^{38}$ erg s$^{-1}$, gives a half-height at a
radius close to the neutron star ($1.01\cdot R_{\rm {*}}$)
varying between $\sim $ 25--50 m.

As a first approximation, in Fig. 4 (left), we show the variation of
{\it h} with $H_{\rm {eq}}$, which was calculated using
the total luminosity of each source to give $\dot M$ {\it via}
$L_{\rm {Tot}}$ = $G M_{\rm {*}} \dot M /R_{\rm {*}}$. It can be seen
that there is agreement between $h$ and $H_{\rm {eq}}$; however, the
lower luminosity points fall below the line $h$ = $H_{\rm {eq}}$. This would be
expected, since for lower luminosities $H_{\rm {eq}}$ is never
achieved since the disk is radiatively-supported over an insufficient
radial extent.

We calculate the actual maximum height of the disk in such cases as follows. The disk 
continues to be thick while the radiation pressure $p_{\rm {r}} >>
p_{\rm {g}}$, 
the gas pressure. Czerny \& Elvis (1987) discuss the transition from a thin
disk to a thick disk. From their Eqn. 19, it can be seen that the
disk height becomes 90\% of $H_{\rm {eq}}$ for 
$p_{\rm {r}}/(p_{\rm {r}}\,+\,p_{\rm {g}})$ = 10.
For a calculation of disk height accurate to $\sim $10\%, we have derived the radius
$r_{\rm {10}}$ at which $p_{\rm {r}}$=10 $p_{\rm {g}}$, using the expressions for radiation pressure
and gas pressure given by Shakura \& Sunyaev (1976), to be 
$r_{\rm {10}}$ = $10.1\, M_{\rm {1}}^{\rm {-3/21}} \alpha ^{\rm {2/21}}
\dot M_{\rm {16}}^{\rm {16/21}} f^{\rm {4/21}}$ km, where $M_{\rm {1}}$ 
is the mass of the neutron star in Solar masses, $\alpha $
is the viscosity parameter, $\dot M_{\rm {16}}$ is in units of
10$^{16}$ g s$^{-1}$ and $f$ = $[1 - (R_{\rm {*}}/r)^{\rm {1/2}}]^{\rm {1/4}}$. 
This equation can be solved numerically for $r_{\rm {10}}$ for particular values of
$\dot M$. For example, using $\dot M$ values 
of $\rm {10^{17}}$ and $\rm {10^{18}}$ g s$^{-1}$ equivalent to total 
luminosities of $\rm {2\times 10^{37}}$ and $\rm {2\times 10^{38}}$ erg
s$^{-1}$, and assuming $\alpha$ = 0.1, $r_{\rm {10}}$ is $\sim $50 km and $\sim $300 km respectively,
but for $\rm {3\times 10^{36}}$ erg s$^{-1}$, $r_{\rm {10}}$ = 10.4 km only.
The actual maximum height of the disk will be $H(r_{\rm {10}})$.
In Fig 4 (right) we show the variation of $h$ with $H(r_{\rm {10}})$,
the more realistic calculation of disk height. 
It can be seen that there is remarkably good agreement between the height of the emission
region and the height of the radiatively-supported disk, most points
lying close to the line {\it h = H}.

In the case of GX\th 5-1, the agreement with {\it h = H} is poor, as
{\it h} = 110
km and {\it H} = 43 km. However the blackbody radius in this case is 33 km
showing that the emission is from a sphere three times larger than the
star. There is a clear implication that the accretion flow is so strong
that a spherical cloud of matter around the star has formed which
is responsible for the blackbody emission. If we compare {\it H} with the
height of this sphere, i.e. $R_{\rm {BB}}$ = 33 km, there
is much better agreement between these quantities, again
suggesting that the height of the disk determines the height of the
emission region. In Fig. 4 we plot triangles
for GX\th 5-1 and the other very bright source GX\th 13+1 to show
their expected positions at {\it h} = $R_{\rm {BB}}$. 
Overall, there is good agreement between {\it h} and {\it H} in Fig. 4
(right) over about 3 orders of magnitude on each axis which is strong evidence for
the significance of the agreement. Appropriate errors are shown for all
points derived from 90\% confidence limits in spectra parameters.
Error in source distance will affect both {\it h} and {\it H};
adopting 10 km for the stellar radius will affect {\it h},
a value of 12 km reducing heights by 20\%. It is also clear that we
may introduce error by using the luminosity in the band 1--30 keV for 
calculating {\it H}, but this range was chosen to avoid extrapolating {\it ASCA}
results too far above 10 keV. Using our work on XB\th 1916-053 from
\sax\ (Church et al. 1998a) we have estimated possible errors.
For this source, with $E_{\rm {CO}}$ = 80 keV, and the total
luminosity would have been underestimated by 28\% using the band 1--30 keV. 
For $E_{\rm {CO}}$ = 20 keV, the error falls to 8\%
and for $E_{\rm {CO}}$ = 10 keV, the error is 1\%. Thus there can
be a further source of error in {\it H} requiring points to be moved to 
the right by up to 30\%. Clearly, broadband spectral analysis is desirable for all sources in
the survey. The implications of Fig. 4 will be discussed in the next section.

\section{Discussion}

We have shown that the two-component blackbody plus Comptonization
model gives a good description of the {\it ASCA} 
spectra of the Atoll and Z-track sources analysed in this survey. We thus now propose 
that this model suggested by us as a unifying model for the dipping sources 
(Church \& Ba\l uci\'nska-Church 1995) gives a good description of LMXB in general.
Fitting results for the disk blackbody plus
Comptonization model were in 8 out of 13 cases unphysical having
values of inner accretion disk radius substantially smaller than the
neutron star radius making it unlikely that the blackbody emission
originates in the disk. Our previous results for 
the dipping sources in particular (see Sect. 1), prove that the Comptonizing region
is extended, indicating a flat, vertically shallow ADC above the inner accretion 
disk. Thus we do not expect significant Comptonization of blackbody emission 
from the neutron star in such an ADC since it is geometrically unlikely that 
blackbody photons will pass through this region. It is similarly
unlikely that substantial Comptonization of the neutron star emission takes place
in the inner radiatively-supported disk as this would require
long path lengths of photons through the disk.

Next, we have shown that use of the two-component model reveals a broad dependence 
of the blackbody luminosity on the total luminosity not previously known for LMXB in general, 
the blackbody luminosity falling more rapidly than the total luminosity as the mass accretion rate 
decreases. The major change in blackbody emission has been shown to result primarily from changes 
in the emitting area, not the temperature. Assuming that 
the emission originates in an equatorial belt on the neutron star
having radius $\sim $10 km, we have shown 
that its height agrees well with the height of the radiatively-supported inner accretion 
disk. This agreement extends to the less luminous sources where the radiatively-supported 
disk does not achieve its equilibrium height, and there is good agreement between {\it h} 
and $H$ calculated at the position where $p_{\rm {r}}$ = $10\,p_{\rm {g}}$. 

Possible explanations of the result that {\it h = H} are of two types; those in which the 
accretion disk height {\it directly} determines $H$, and those in which it does not. The 
direct mechanism requires radial (advective) flow of material between the inner disk edge 
and the star. Although there has been a very large amount of theoretical work on advective 
flow in black hole binaries (e.g. Abramowicz et al. 1996, and references therein), little 
work has been carried out for LMXB. It is not known whether the sonic point lies within the 
inner disk or within the star. However, Popham \& Sunyaev (2000) show
that the radial velocity increases by two orders of magnitude in the
inner disk, but it is not known whether accretion flow can cross the gap
between the inner disk edge and the star. Resolving this question will require
detailed two- or three-dimensional hydrodynamic modelling.

Inogamov \& Sunyaev (1999) have recently shown that accretion flow meeting the surface of 
the neutron star in the equatorial plane will spread over the surface of the star 
producing regions of enhanced X-ray brightness of vertical extent increasing with luminosity, as 
is the case in our observational result. The mechanism proposed is independent of the 
accretion disk height and so the {\it h = H} agreement would result
from {\it H} being a measure of {\it L}. Preliminary work shows
encouraging agreement between the emitting height on the neutron star
predicted by Inogamov and Sunyaev.
A detailed comparison between the present results and this theory
will be made in a further paper. 

The above results have a bearing on alternative models for LMXB. Our results are very inconsistent with 
models in which the blackbody originates in the accretion disk, and
the agreement of {\it h = H} is positive 
evidence for neutron star blackbody emission. Additionally, we have  used the timescales for 
dip ingress and egress in several observations of dipping sources
(Church 2000) to obtain 
the radius of the extended Comptonizing ADC region: $r_{\rm {ADC}}$, typically $\rm {5\times 10^9}$ cm. 
These measurements are supported by the fact that the Comptonized emission component is gradually covered 
during dipping. Thus the ADC is 5,000 times larger than the neutron
star, and this rules out models in which Comptonization is supposed to take place in the
neighbourhood of the neutron star.

We clearly need to improve our knowledge of the emission parameters of some of the sources studied by 
obtaining broadband spectra extending to $\sim $100 keV, allowing the cut-off energy to be measured, 
which has not in general been possible in the present work. Further investigation of both very bright 
sources and very weak sources will also be of interest. A complete description of the 
blackbody and Comptonized emission in LMXB must be more complicated
than given here. For example, during flaring, major changes take place
on the neutron star and inner disk and this is investigated in work
currently taking place (Ba\l uci\'nska-Church et al. 2001).

\begin{acknowledgements}
MJC and MBC thank the Royal Society and the British Council for financial support, and 
Andrew King, W\l odek Klu\'zniak, Nail Inogamov and Rashid Sunyaev for 
helpful discussions.
\end{acknowledgements}

\end{document}